\documentclass{article}

\usepackage{epsfig}
\usepackage{graphics}
\usepackage{graphicx}
\usepackage[centertags]{amsmath}
\usepackage{amsfonts}
\usepackage{amssymb}
\usepackage{url}
\usepackage{subfigure}
\usepackage{hyperref}

\begin{document}

\title{A Note on Elementary Cellular Automata Classification}

\author{Genaro J. Mart\'{\i}nez}

\date{June 17, 2013\footnote{Accepted by publish in {\it Journal of Cellular Automata}. \url{http://www.oldcitypublishing.com/JCA/JCA.html}}}

\maketitle

\begin{centering}
$^1$ Departamento de Ciencias e Ingenier{\'i}a de la Computaci{\'o}n, \\ Escuela Superior de C\'omputo, Instituto Polit\'ecnico Nacional, M\'exico, D.F. \\
$^2$ International Centre of Unconventional Computing, \\ University of the West of England, BS16 1QY Bristol, United Kingdom \\
\url{genaro.martinez@uwe.ac.uk} \\

\end{centering}

\begin{abstract}
We overview and compare classifications of elementary cellular automata, including Wolfram's, Wuensche's, Li and Packard, communication complexity, power spectral, topological, surface, compression, lattices, and morphological diversity classifications. This paper summarises several classifications of elementary cellular automata (ECA) and compares them with a newly proposed one, that induced by endowing rules with memory.

\end{abstract}

\tableofcontents

\section{Preamble}

This paper describes several ECA classifications. Consequently, we can compare quickly some basic properties and use these relations for future classifications.

In this respect, a remarkable and important result was published by Culik II and Yu in 1988, who proved that such classifications are undecidable \cite{kn:CY88}. However, novel and recombined techniques continue reporting more approaches \cite{kn:Wue99}. Also, we will compare each classification with a recent one, {\it memory classification}.

Our motivation begins with a classification of ECA rules, but composed with memory functions (for details about cellular automata with memory (CAM), please see \cite{kn:Alo09, kn:Alo11}).

We know from previous studies that ECA composed with memory (ECAM) yield another automaton with uniform, periodic, chaotic, or complex behaviour. In particular, previous analysis has shown that chaotic ECA rules composed with memory are capable of displaying complex behaviour (for ECA rule 30 see \cite{kn:MAA10}, for ECA rule 45 see \cite{kn:MAA12}, and for ECA rule 126 see \cite{kn:MAS10}). Of course, the composition produces a new rule, but with elements of the original ECA rule. This way, memory functions help to `discover' hidden information in dynamical systems from simple functions (or rules), and ``transform'' simple and chaotic rules to complex rules or vice versa \cite{kn:MAA}. Another approach to get complex rules was displayed by Gunji's analysis, deriving complex rules from other CA rules, using intermediate layer lattices \cite{kn:Gunji10}.

\newpage

\begin{itemize}
\item[] {\bf Class I}: leads to uniform behaviour, \\

\begin{figure}[th]
\centerline{\includegraphics[width=4.6in]{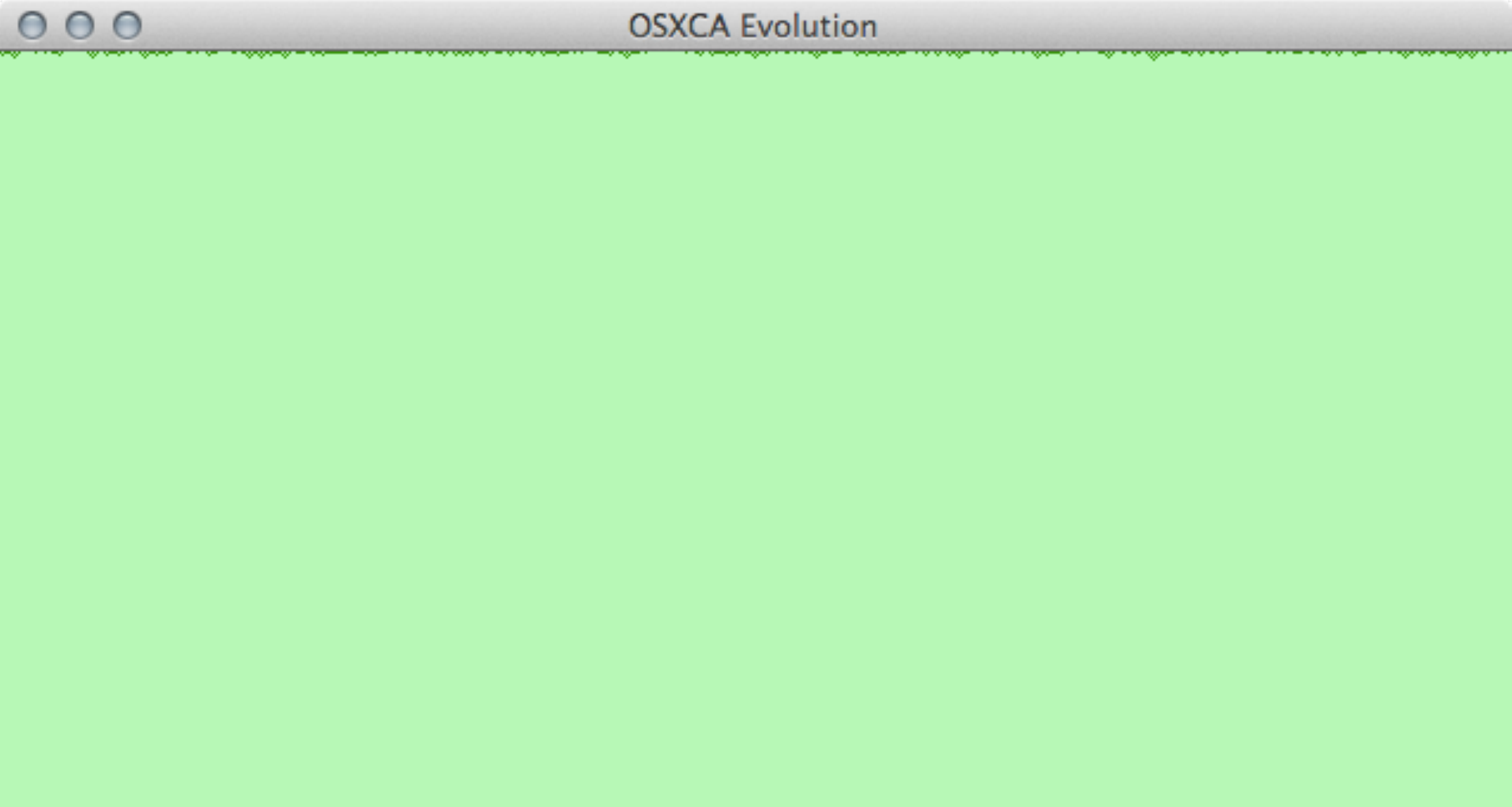}}
\caption{ECA rule 32.}
\label{rule32}
\end{figure}
\end{itemize}

\begin{itemize}
\item[] {\bf Class II}: leads to periodic behaviour, \\

\begin{figure}[th]
\centerline{\includegraphics[width=4.6in]{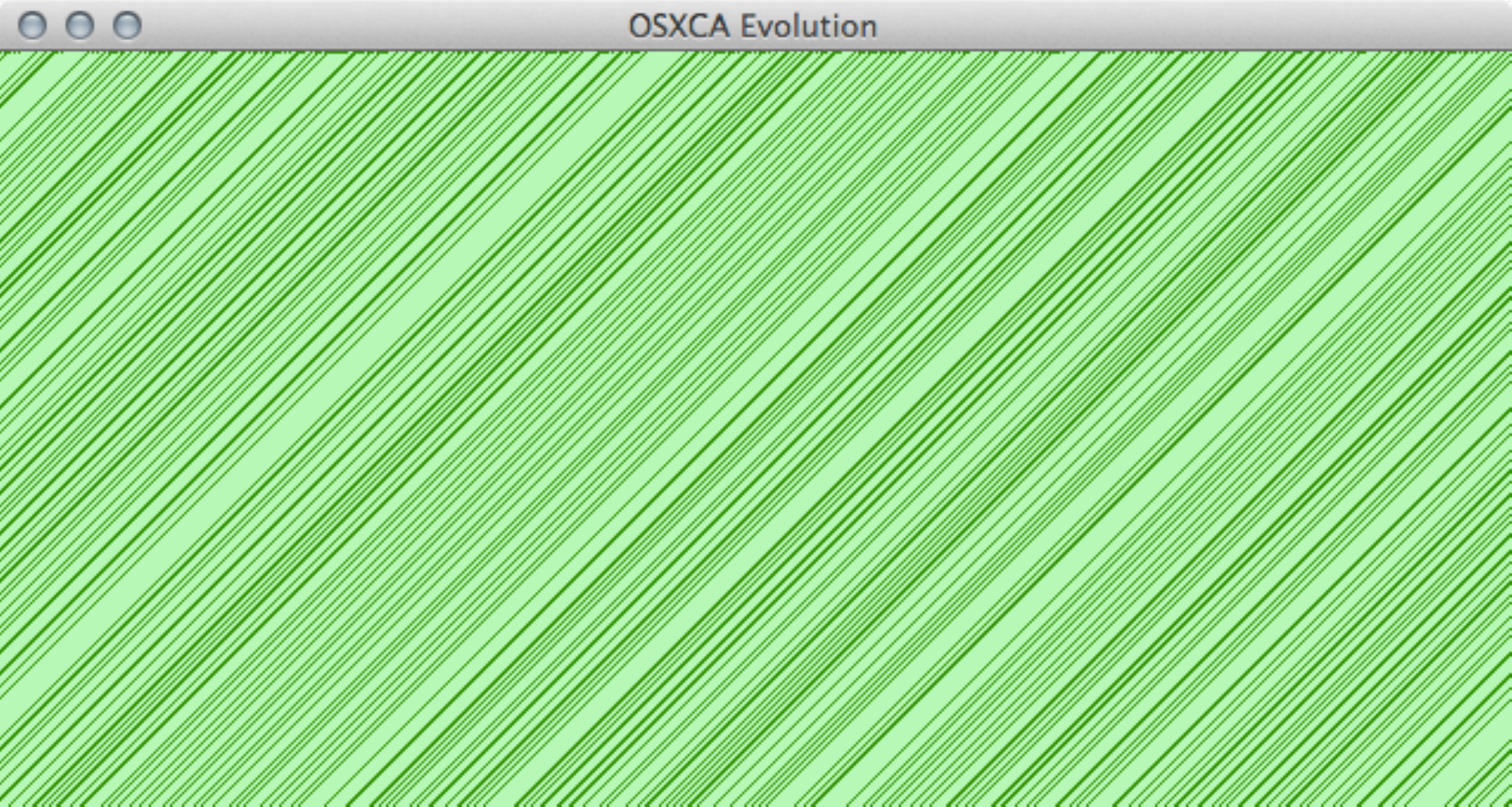}}
\caption{ECA rule 10.}
\label{rule10}
\end{figure}
\end{itemize}

\newpage

\begin{itemize}
\item[] {\bf Class III:} leads to chaotic behaviour, \\

\begin{figure}[th]
\centerline{\includegraphics[width=4.6in]{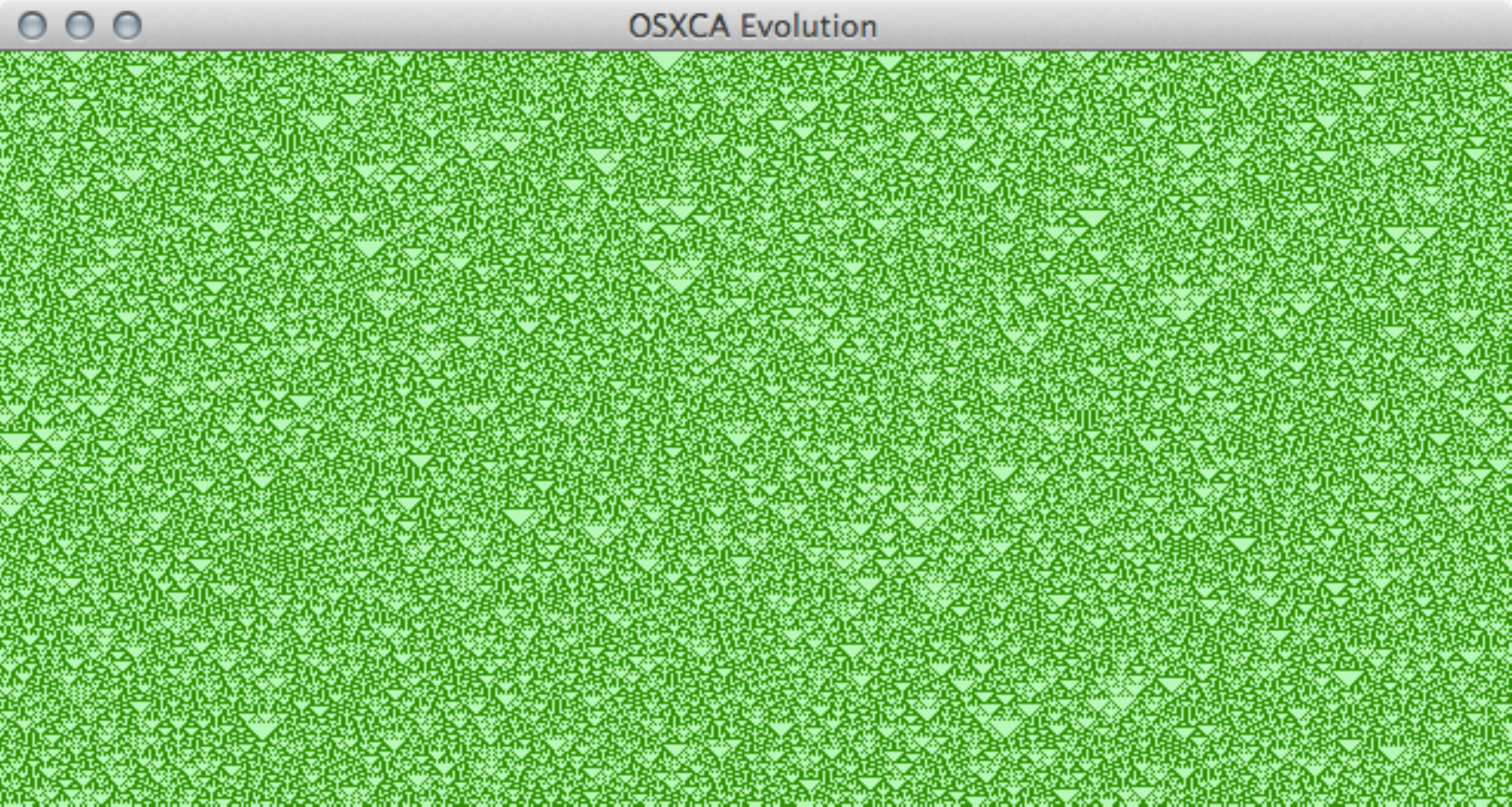}}
\caption{ECA rule 90.}
\label{rule90}
\end{figure}
\end{itemize}

\begin{itemize}
\item[] {\bf Class IV}: leads to complex behaviour. \\

\begin{figure}[th]
\centerline{\includegraphics[width=4.6in]{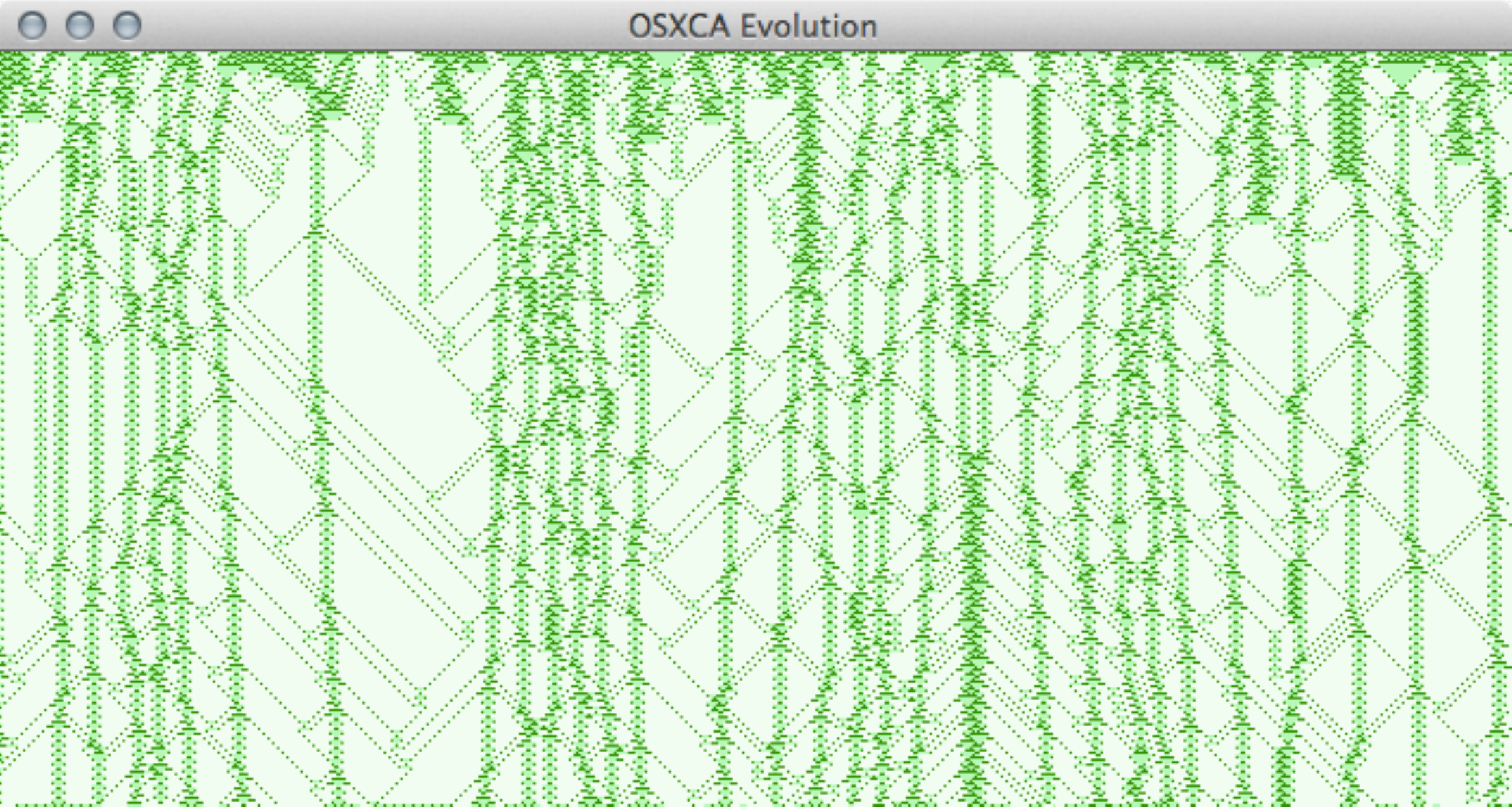}}
\caption{ECA rule 54 (filtered).}
\label{rule54}
\end{figure}
\end{itemize}


\begin{table}
\small
\centering
\begin{tabular}{|c|l|c|l|}
  \hline
  rule & equivalent rules & rule & equivalent rules \\
  \hline
      0 & 255 & 56 & 98, 185, 227 \\
      1 & 127 & 57 & 99 \\
      2 & 16, 191, 247 & 58 & 114, 163, 177 \\
      3 & 17, 63, 119 & 60 & 102, 153, 195 \\
      4 & 223 & 62 & 118, 131, 145 \\
      5 & 95 & 72 & 237 \\
      6 & 20, 159, 215 & 73 & 109 \\
      7 & 21, 31, 87 & 74 & 88, 173, 229 \\
      8 & 64, 239, 253 & 76 & 205 \\
      9 & 65, 111, 125 & 77 & - \\
    10 & 80, 175, 245 & 78 & 92, 141, 197 \\
    11 & 47, 81, 117 & 90 & 165 \\
    12 & 68, 207, 221 & 94 & 133 \\
    13 & 69, 79, 93 & 104 & 233 \\
    14 & 84, 143, 213 & 105 & - \\
    15 & 85 & 106 & 120, 169, 225 \\
    18 & 183 & 108 & 201 \\
    19 & 55 & 110 & 124, 137, 193 \\
    22 & 151 & 122 & 161 \\
    23 & - & 126 & 129 \\
    24 & 66, 189, 231 & 128 & 254 \\
    25 & 61, 67, 103 & 130 & 144, 190, 246 \\
    26 & 82, 167, 181 & 132 & 222 \\
    27 & 39, 53, 83 & 134 & 148, 158, 214 \\
    28 & 70, 157, 199 & 136 & 192, 238, 252 \\
    29 & 71 & 138 & 174, 208, 224 \\
    30 & 86, 135, 149 & 140 & 196, 206, 220 \\
    32 & 251 & 142 & 212 \\
    33 & 123 & 146 & 182 \\
    34 & 48, 187, 243 & 150 & - \\
    35 & 49, 59, 115 & 152 & 188, 194, 230 \\
    36 & 219 & 154 & 166, 180, 210 \\
    37 & 91 & 156 & 198 \\
    38 & 52, 155, 211 & 160 & 250 \\
    40 & 96, 235, 249 & 162 & 176, 186, 242 \\
    41 & 97, 107, 121 & 164 & 218 \\
    42 & 112, 171, 241 & 168 & 224, 234, 248 \\
    43 & 113 & 170 & 240 \\
    44 & 100, 203, 217 & 172 & 202, 216, 228 \\
    45 & 75, 89, 101 & 178 & - \\
    46 & 116, 139, 209 & 184 & 226 \\
    50 & 179 & 200 & 236 \\
    51 & - & 204 & - \\
    54 & 147 & 232 & - \\
    \hline
\end{tabular}
\caption{88 equivalent ECA rules.}
\label{eqv_rules}
\end{table}


An ECA is the most basic one-dimensional CA representation. ECA have two states in its alphabet $\Sigma=\{0,1\}$ and two close neighbours with respect to one central cell. Thus, a central cell $x_i$ may take an element of the alphabet and update simultaneously on a configuration, this update is done accordingly to an evolution rule $\varphi$. Hence, we have that an evolution rule can be expressed as\,: $\varphi(x_{i-r}^{t}, \ldots, x_{i}^{t}, \ldots, x_{x+r}^{t}) \rightarrow x_{i}^{t+1}$, where $r$ represents the number of left and right neighbours, and $x \in \Sigma$. This way, the evolution function in ECA domain is $\varphi(x_{i-1}^{t}, x_{i}^{t}, x_{x+1}^{t}) \rightarrow x_{i}^{t+1}$. If $k=|\Sigma|$ then we can calculate the number of neighbours and evolution rules. $k^{2r+1}$ determines the number of neighbours and $k^{k^{2r+1}}$ the number of different evolution rules. Therefore, for ECA domain $(k=2, r=1)$ there are 256 different evolution rules \cite{kn:Wolf94}.

In 1983, Wolfram establishes a classification in CA \cite{kn:Wolf83}, with four classes that describe global behaviour from random initial conditions. Figures~\ref{rule32}--\ref{rule54} illustrate each class with a particular evolution rule. All evolutions evolve on a ring of 637 cells to 318 generations, using the same random initial condition at 50\% of cells in state zero or one.

Some researchers found that CA rules have equivalences. The earliest was Walker and Ashby in 1966 \cite{kn:WA66}. Next was Martin, Odlyzko, and Wolfram in 1984 \cite{kn:MOW84}

Li and Packard also found that ECA rules have equivalences \cite{kn:LP90}. Consequently, a similar behaviour can be projected and organised in set of rules, thus ECA rule-space has 88 equivalent sets of rules, as shown in table~\ref{eqv_rules}. Recently such a property was restudied and obtained across topological properties in \cite{kn:GST07}.

\section{The bases}

This study considers two main analyses: That of Wolfram and Wuensche's classifications.

\subsection{Wolfram's classification}

In his seminal book ``Theory and Applications of Cellular Automata'', Wolfram had included an extended number of properties in one-dimensional CA, displayed in ``Tables of Cellular Automata Properties'' section \cite{kn:Wolf86}. Here we have reviewed and updated such a classification with the {\it WolframAlpha} engine.\footnote{WolframAlpha computational knowledge engine \url{http://www.wolframalpha.com/}.}

Wolfram's classification is shown in Table~\ref{Wolfram_classes}. This classification establishes that from random initial conditions a CA will reach in a long time, one of the following classes of global behaviour.

\begin{itemize}
\item[] {\bf Class I}: evolve to {\it uniform} behaviour;
\item[] {\bf Class II}: evolve to {\it periodic} behaviour;
\item[] {\bf Class III}: evolve to {\it chaotic} behaviour;
\item[] {\bf Class IV}: evolve to {\it complex} behaviour.
\end{itemize}





\begin{table}[th]
\centering
\begin{tabular}{lcl}
  \hline\noalign{\smallskip}
  \multicolumn{3}{p{2cm}}{classification} \\
  \cline{1-2}\noalign{\smallskip}
  type & num. & rules \\
  \noalign{\smallskip}\hline\noalign{\smallskip}
  {\sf class I} & 8 & 0, 8, 32, 40, 128, 136, 160, 168. \\
  {\sf class II} & 65 & 1, 2, 3, 4, 5, 6, 7, 9, 10, 11, 12, 13,  14, 15, 19, 23, \\
               &      & 24, 25, 26, 27, 28, 29, 33, 34, 35, 36, 37, 38, 42, \\
               &      & 43, 44, 46, 50, 51, 56, 57, 58, 62, 72, 73, 74, 76, \\
               &      & 77, 78, 94, 104, 108, 130, 132, 134, 138, 140, 142, \\
               &      & 152, 154, 156, 162, 164, 170, 172, 178, 184, 200, \\
               &      & 204, 232. \\
  {\sf class III} & 11 & 18, 22, 30, 45, 60, 90, 105, 122, 126, 146, 150. \\
  {\sf class IV} & 4 & 41, 54, 106, 110. \\
  \noalign{\smallskip}\hline
\end{tabular}
\caption{Wolfram's classification relation.}
\label{Wolfram_classes}
\end{table}

The main interest of {\it chaotic} rules relate to developing cryptography, random number generators, and fields of attraction. However, the so called class IV or {\it complex} rules have captured most attention given their potential for computational universality, and their applications in artificial life by the simulations of particles, waves, mobile self localizations, or gliders.\footnote{Complex Cellular Automata Repository \url{http://uncomp.uwe.ac.uk/genaro/Complex_CA_repository.html}.} Their capacity to contain intrinsically {\it complex systems} \cite{kn:Wolf02, kn:Wolf84, kn:Kau93, kn:Ada02, kn:TM87, kn:Lang86}. This kind of discrepancy between chaotic rules, and complex rules capable of computational universality, are discussed in the CA literature (for details please see \cite{kn:MSZ13a, kn:MSZ13b}).

\subsection{Wuensche's equivalences}

Wuensche and Lesser did a detailed analysis on basin of attraction fields and rule clusters of equivalences in ECA, published in their book ``The Global Dynamics of Cellular Automata'' written in 1992 \cite{kn:WL92}.

In this book, Wuensche recognises three main transformations that relate equivalent ECA rules expressed in {\it rule clusters}. Such transformations are: reflection, negation, and complementation. Moreover, these equivalences can be explored quickly and automatically for other CA with free software {\it Discrete Dynamics Lab}\footnote{Discrete Dynamics Lab (DDLab) \url{http://www.ddlab.org/}.} \cite{kn:Wue11}.

Rule clusters classification displays 88 equivalences (as is shown in Table~\ref{eqv_rules}) in ECA, and besides establishes basically three kind of  symmetries: symmetric, semi-asymmetric, and full-asymmetric set of rules \cite{kn:WL92}.




\begin{table}[th]
\centering
\begin{tabular}{lcl}
  \hline\noalign{\smallskip}
  \multicolumn{3}{p{2cm}}{equivalences} \\
  \cline{1-2}\noalign{\smallskip}
  type & num. & rules \\
  \noalign{\smallskip}\hline\noalign{\smallskip}
  {\sf symmetric} & 36 & 0, 1, 4, 5, 18, 19, 22, 23, 32, 33, 36, 37, \\
                   &      & 50, 51, 54, 72, 73, 76, 77, 90, 94, 104, \\
                   &      & 105, 108, 122, 126, 128, 132, 146, 150, \\
                   &      & 160, 164, 178, 200, 204, 232. \\
  {\sf semi-asymmetric} & 32 & 2, 3, 6, 7, 8, 9, 12, 13, 26, 27, 30, 34, 35, \\
                   &      & 38, 40, 41, 44, 45, 58, 62, 74, 78, 106, \\
                   &      & 110, 130, 134, 136, 140, 154, 162, 168, \\
                   &      & 172. \\
  {\sf full-asymmetric} & 20 & 10, 11, 14, 15, 24, 25, 28, 29, 42, 43, 46,\\
                   &      & 57, 60, 138, 142, 152, 156, 170, 184. \\
  \noalign{\smallskip}\hline
\end{tabular}
\caption{Wuensche's equivalences relation.}
\label{Wuensche_classes}
\end{table}

Also, these equivalences become refined in that some transformations (refection, negation, or complement) produce identity or another transformation, causing a rule cluster of 4 equivalent rules and their 4 equivalent compliments, to collapse. Where reflection produces the same rule, the rule is called symmetric (otherwise asymmetric) -- but the reflection algorithm allows for semi-asymmetric and fully-asymmetric rules.

\section{ECA with memory (ECAM)}
In this section, we propose a classification based in memory functions. The full analysis of this classification can be found in \cite{kn:MAA}. Here, we will merely compare this classification with the other ones.

\subsection{ECAM's classification}

Memory classification on ECA is investigated in ``Designing Complex Dynamics with Memory'' \cite{kn:MAA}. A ECAM is a ECA composed with a memory function, the new rule open a new and extended domain of rules based in the ECA domain \cite{kn:MAA10}.

Basically, if you select an ECA rule and compose this rule with a memory function (in our analysis we have considered three basic functions\,: majority, minority, and parity). Hence, we will achieve a new rule with a base of a ECA rule. Of course, a number of features from the original ECA rule will be more evident on its ECAM generalisation. Therefore, the memory function will determine if the original ECA rule preserves the same class (respect to Wolfram's classes) or if it changes to another class.

Following this simple principle, we know now that ECA rules composed with memory can be classified in three classes:

\begin{itemize}
\item[] {\bf \em strong}, because the memory functions are unable to transform one class to another;
\item[] {\bf \em moderate}, because the memory function can transform the rule to another class and conserve the same class as well;
\item[] {\bf \em weak}, because the memory functions do most transformations and the rule changes to another different class quickly.
\end{itemize}

This way, the next table displays the ECA classification based in memory functions.




\begin{table}[th]
\centering
\begin{tabular}{lcl}
  \hline\noalign{\smallskip}
  \multicolumn{3}{p{2cm}}{classification} \\
  \cline{1-2}\noalign{\smallskip}
  type & num. & rules \\
  \noalign{\smallskip}\hline\noalign{\smallskip}
  {\sf strong} & 39 & 2, 7, 9, 10, 11, 15, 18, 22, 24, 25, 26, 30, 34, \\
                   &      & 35, 41, 42, 45, 46, 54, 56, 57, 58, 62, 94, 106, \\
                   &      & 108, 110, 122, 126, 128, 130, 138, 146, 152, \\
                   &      & 154, 162, 170, 178, 184. \\
  {\sf moderate} & 34 & 1, 3, 4, 5, 6, 8, 13, 14, 27, 28, 29, 32, 33, 37, \\
                   &      & 38, 40, 43, 44, 72, 73, 74, 77, 78, 104, 132, \\
                   &      & 134, 136, 140, 142, 156, 160, 164, 168, 172. \\
  {\sf weak} & 15 & 0, 12, 19, 23, 36, 50, 51, 60, 76, 90, 105, 150, \\
                   &      & 200, 204, 232. \\
  \noalign{\smallskip}\hline
\end{tabular}
\caption{ECAM's classification relation.}
\label{ECAM_classes}
\end{table}

\subsection{Some relevant properties}

Memory classification presents a number of interesting properties.

We have ECA rules which composed with a particular kind of memory  are able of reach another class including the original dynamic. The main feature is that, at least, this rule with memory is able to reach every different class. Rules with this property are called  {\it universal ECAM} (5 rules).

\begin{center}
    \begin{tabular}{ r p{9cm}}
    {\sf universal ECAM:} & 22, 54, 146, 130, 152.
    \end{tabular}
\end{center}

Particularly, all these UECAM are classified {\sf strong} in ECAM's classification.

\begin{center}
    \begin{tabular}{ r p{9cm}}
    {\sf strong:} & 22, 54, 146, 130, 152. \\
    {\sf moderate:} & - \\
    {\sf weak:} & -
    \end{tabular}
\end{center}

On the other hand, we have  ECA that when composed with memory are able to yield a complex ECAM but with elements of the original ECA rule. They are called {\it complex ECAM} (44 rules).

\begin{center}
    \begin{tabular}{ r p{9cm}}
    {\sf complex ECAM:} & 6, 9, 10, 11, 13, 15, 22, 24, 25, 26, 27, 30, 33, 35, 38, 40, 41, 42, 44, 46, 54, 57, 58, 62, 72, 77, 78, 106, 108, 110, 122, 126, 130, 132, 138, 142, 146, 152, 156, 162, 170, 172, 178, 184.
    \end{tabular}
\end{center}

\noindent and they can be particularised in terms of ECAM's classification, as follows:

\begin{center}
    \begin{tabular}{ r p{9cm}}
    {\sf strong:} & 9, 10, 11, 15, 22, 24, 25, 26, 30, 35, 41, 42, 46, 54, 57, 58, 62, 106, 108, 110, 122, 126, 130, 138, 146, 152, 162, 170, 178, 184. \\
    {\sf moderate:} & 6, 13, 27, 33, 38, 40, 44, 72, 77, 78, 132, 142, 156, 172. \\
    {\sf weak:} & -
    \end{tabular}
\end{center}

It is remarkable that none of the rules classified in {\sf weak} class is able to reach complex behaviour. These set of rules are strongly robust to any perturbation in terms of ECAM's classification.

\section{ECA classifications versus memory classification}
In these sections, we will compare several ECA classifications reported in CA literature all along  the CA-history versus memory classification.

\subsection{Wolfram's classification (1984)}

Wolfram's classification in ``Universality and complexity in cellular automata'', establishes four classes:

\begin{center}
\{{\sf uniform (class I), periodic (class II), chaotic (class III), complex (class IV)}\}
\end{center}

For details please see \cite{kn:Wolf94, kn:Wolf02}.

\begin{center}
    \begin{tabular}{ r p{9cm}}
    {\bf class I:} & 0, 8, 32, 40, 128, 136, 160, 168. \\
    & \\
    {\sf strong:} & 128. \\
    {\sf moderate:} & 8, 32, 40, 136, 160, 168. \\
    {\sf weak:} & 0.
    \end{tabular}
\end{center}

\begin{center}
    \begin{tabular}{ r p{9cm}}
    {\bf class II:} & 1, 2, 3, 4, 5, 6, 7, 9, 10, 11, 12, 13, 14, 15, 19, 23, 24, 25, 26, 27, 28, 29, 33, 34, 35, 36, 37, 38, 42, 43, 44, 46, 50, 51, 56, 57, 58, 62, 72, 73, 74, 76, 77, 78, 94, 104, 108, 130, 132, 134, 138, 140, 142, 152, 154, 156, 162, 164, 170, 172, 178, 184, 200, 204, 232. \\
    & \\
    {\sf strong:} & 2, 7, 9, 10, 11, 15, 24, 25, 26, 34, 35, 42, 46, 56, 57, 58, 62, 94, 108, 130, 138, 152, 154, 162, 170, 178, 184. \\
    {\sf moderate:} & 1, 3, 4, 5, 6, 13, 14, 27, 28, 29, 33, 37, 38, 43, 44, 72, 73, 74, 77, 78, 104, 132, 134, 140, 142, 156, 164, 172. \\
    {\sf weak:} & 12, 19, 23, 36, 50, 51, 76, 200, 204, 232.
    \end{tabular}
\end{center}

\begin{center}
    \begin{tabular}{ r p{9cm}}
    {\bf class III:} & 18, 22, 30, 45, 60, 90, 105, 122, 126, 146, 150. \\
    & \\
    {\sf strong:} & 18, 22, 30, 45, 122, 126, 146. \\
    {\sf moderate:} & - \\
    {\sf weak:} & 60, 90, 105, 150.
    \end{tabular}
\end{center}

\begin{center}
    \begin{tabular}{ r p{9cm}}
    {\bf class IV:} & 41, 54, 106, 110. \\
    & \\
    {\sf strong:} & 41, 54, 106, 110. \\
    {\sf moderate:} & - \\
    {\sf weak:} & -
    \end{tabular}
\end{center}

\subsection{Li and Packard's classification (1990)}

Li and Packard's classification in ``The Structure of the Elementary Cellular Automata Rule Space'', establishes five ECA classes:

\begin{center}
\{{\sf null, fixed point, periodic, locally chaotic, chaotic}\}.
\end{center}

For details please see \cite{kn:LP90}.

\begin{center}
    \begin{tabular}{ r p{9cm}}
    {\bf null:} & 0, 8, 32, 40, 128, 136, 160, 168. \\
    & \\
    {\sf strong:} & 128. \\
    {\sf moderate:} & 8, 32, 40, 136, 160, 168. \\
    {\sf weak:} & 0.
    \end{tabular}
\end{center}

\begin{center}
    \begin{tabular}{ r p{9cm}}
    {\bf fixed point:} & 2, 4, 10, 12, 13, 24, 34, 36, 42, 44, 46, 56, 57, 58, 72, 76, 77, 78, 104, 130, 132, 138, 140, 152, 162, 164, 170, 172, 184, 200, 204, 232. \\
    & \\
    {\sf strong:} & 2, 10, 24, 34, 42, 46, 56, 57, 58, 130, 138, 152, 162, 170, 184. \\
    {\sf moderate:} & 4, 13, 44, 72, 77, 78, 104, 132, 140, 164, 172. \\
    {\sf weak:} & 12, 36, 76, 200, 204, 232.
    \end{tabular}
\end{center}

\begin{center}
    \begin{tabular}{ r p{9cm}}
    {\bf periodic:} & 1, 3, 5, 6, 7, 9, 11, 14, 15, 19, 23, 25, 27, 28, 29, 33, 35, 37, 38, 41, 43, 50, 51, 74, 94, 108, 131(62), 134, 142, 156, 178. \\
    & \\
    {\sf strong:} & 7, 9, 11, 15, 25, 35, 41, 62, 94, 108, 178. \\
    {\sf moderate:} & 1, 3, 5, 6, 14, 27, 28, 29, 33, 37, 38, 43, 74, 134, 142, 156. \\
    {\sf weak:} & 19, 23, 50, 51.
    \end{tabular}
\end{center}

\begin{center}
    \begin{tabular}{ r p{9cm}}
    {\bf locally chaotic:} & 26, 73, 154. \\
    & \\
    {\sf strong:} & 26, 154. \\
    {\sf moderate:} & 73. \\
    {\sf weak:} & -
    \end{tabular}
\end{center}

\begin{center}
    \begin{tabular}{ r p{9cm}}
    {\bf chaotic:} & 18, 22, 30, 45, 54, 60, 90, 105, 106, 132, 129(126), 137(110), 146, 150, 161(122). \\
    & \\
    {\sf strong:} & 18, 22, 30, 45, 54, 106, 122, 126, 110, 122, 146. \\
    {\sf moderate:} & - \\
    {\sf weak:} & 60, 90, 105, 150.
    \end{tabular}
\end{center}

\subsection{Wuensche's equivalences (1992)}

Wuensche's equivalences in ``The Global Dynamics of Cellular Automata'', establishes three ECA kinds of symmetries:

\begin{center}
\{{\sf symmetric, semi-asymmetric, full-asymmetric}\}.
\end{center}

For details please see \cite{kn:WL92}.

\begin{center}
    \begin{tabular}{ r p{9cm}}
    {\bf symmetric:} & 0, 1, 4, 5, 18, 19, 22, 23, 32, 33, 36, 37, 50, 51, 54, 72, 73, 76, 77, 90, 94, 104, 105, 108, 122, 126, 128, 132, 146, 150, 160, 164, 178, 200, 204, 232. \\
    & \\
    {\sf strong:} & 18, 22, 54, 108, 122, 126, 128, 146, 178. \\
    {\sf moderate:} & 1, 4, 5, 32, 33, 72, 73, 77, 104, 132, 160, 164. \\
    {\sf weak:} & 0, 19, 23, 36, 50, 51, 76, 90, 105, 150, 200, 204.
    \end{tabular}
\end{center}

\begin{center}
    \begin{tabular}{ r p{8cm}}
    {\bf semi-asymmetric:} & 2, 3, 6, 7, 8, 9, 12, 13, 26, 27, 30, 34, 35, 38, 40, 41, 44, 45, 58, 62, 74, 78, 106, 110, 130, 134, 136, 140, 154, 162, 168, 172. \\
    & \\
    {\sf strong:} & 2, 7, 9, 26, 30, 34, 35, 41, 45, 58, 62, 106, 110, 130, 154, 162. \\
    {\sf moderate:} & 3, 6, 8, 13, 27, 38, 40, 44, 74, 78, 134, 136, 140, 168, 172. \\
    {\sf weak:} & 12.
    \end{tabular}
\end{center}

\begin{center}
    \begin{tabular}{ r p{8cm}}
    {\bf full-asymmetric:} & 10, 11, 14, 15, 24, 25, 28, 29, 42, 43, 46, 57, 60, 138, 142, 152, 156, 170, 184. \\
    & \\
    {\sf strong:} & 10, 11, 15, 24, 25, 42, 46, 57, 138, 152, 170, 184. \\
    {\sf moderate:} & 14, 28, 29, 43, 142, 156. \\
    {\sf weak:} & 60.
    \end{tabular}
\end{center}

Also, Wuensche establishes a set of ``maximally chaotic'' rules or known as ``chain rules'' (for details please see \cite{kn:Wue09}).

\begin{center}
    \begin{tabular}{ r p{9cm}}
    {\bf chain rules:} & 30, 45, 106, 154. \\
    & \\
    {\sf strong:} & 30, 45, 106, 154. \\
    {\sf moderate:} & - \\
    {\sf weak:} & -
    \end{tabular}
\end{center}

\subsection{Index complexity classification (2002)}

Index complexity in ``A Nonlinear Dynamics Perspective of WolframÕs New Kind of Science. Part I: Threshold of Complexity'', establishes three ECA classes:

\begin{center}
\{{\sf red ($k=1$), blue ($k=2$), green ($k=3$)}\}.
\end{center}

For details please see \cite{kn:CYD02}.

\begin{center}
    \begin{tabular}{ r p{9cm}}
    {\bf red ($k=1$):} & 0, 1, 2, 3, 4, 5, 7, 8, 10, 11, 12, 13, 14, 15, 19, 23, 32, 34, 35, 42, 43, 50, 51, 76, 77, 128, 136, 138, 140, 142, 160, 162, 168, 170, 178, 200, 204, 232. \\
    & \\
    {\sf strong:} & 2, 7, 10, 11, 15, 34, 35, 42, 128, 138, 162, 170, 178. \\
    {\sf moderate:} & 1, 3, 4, 5, 8, 13, 14, 32, 43, 77, 136, 140, 142, 160, 168. \\
    {\sf weak:} & 0, 12, 19, 23, 50, 51, 76, 200, 204, 232.
    \end{tabular}
\end{center}

\begin{center}
    \begin{tabular}{ r p{9cm}}
    {\bf blue ($k=2$):} & 6, 9, 18, 22, 24, 25, 26, 28, 30, 33, 36, 37, 38, 40, 41, 44, 45, 54, 56, 57, 60, 62, 72, 73, 74, 90, 94, 104, 106, 108, 110, 122, 126, 130, 132, 134, 146, 152, 154, 156, 164. \\
    & \\
    {\sf strong:} & 9, 18, 22, 24, 25, 26, 30, 41, 45, 54, 56, 57, 62, 94, 106, 108, 110, 122, 126, 130, 146, 152, 154. \\
    {\sf moderate:} & 6, 28, 33, 37, 38, 40, 44, 72, 73, 74, 104, 132, 134, 156, 164. \\
    {\sf weak:} & 36, 60, 90.
    \end{tabular}
\end{center}

\begin{center}
    \begin{tabular}{ r p{9cm}}
    {\bf green ($k=3$):} & 27, 29, 46, 58, 78, 105, 150, 172, 184. \\
    & \\
    {\sf strong:} & 46, 58, 184. \\
    {\sf moderate:} & 27, 29, 78, 172. \\
    {\sf weak:} & 105, 150.
    \end{tabular}
\end{center}

\subsection{Density parameter with $d$-spectrum classification (2003)}

Density parameter with $d$-spectrum in ``Experimental Study of Elementary Cellular Automata Dynamics Using the Density Parameter'', establishes three ECA classes:

\begin{center}
\{{\sf P, H, C}\}.
\end{center}

For details please see \cite{kn:Fat03}.

\begin{center}
    \begin{tabular}{ r p{9cm}}
    {\bf P:} & 0, 1, 2, 3, 4, 5, 6, 7, 8, 9, 10, 11, 12, 13, 14, 15, 19, 23, 24, 25, 27, 28, 29, 32, 33, 34, 35, 36, 37, 38, 40, 42, 43, 44, 50, 51, 56, 57, 58, 62, 72, 74, 76, 77, 78, 104, 108, 128, 130, 132, 134, 136, 138, 140, 142, 152, 156, 160, 162, 164, 168, 170, 172, 178, 184, 200, 204, 232. \\
    & \\
    {\sf strong:} & 2, 7, 9, 10, 11, 15, 24, 25, 34, 35, 42, 56, 57, 58, 62, 108, 128, 130, 138, 152, 162, 170, 178, 184. \\
    {\sf moderate:} & 1, 3, 4, 5, 6, 8, 13, 14, 27, 28, 29, 32, 33, 37, 38, 40, 43, 44, 72, 74, 77, 78, 104, 132, 134, 136, 140, 142, 156, 160, 164, 168, 172. \\
    {\sf weak:} & 0, 12, 19, 23, 36, 50, 51, 76, 200, 204, 232.
    \end{tabular}
\end{center}

\begin{center}
    \begin{tabular}{ r p{9cm}}
    {\bf H:} & 26, 41, 54, 73, 94, 110, 154. \\
    & \\
    {\sf strong:} & 26, 41, 94, 110, 154. \\
    {\sf moderate:} & 73. \\
    {\sf weak:} & -
    \end{tabular}
\end{center}

\begin{center}
    \begin{tabular}{ r p{9cm}}
    {\bf C:} & 18, 22, 30, 45, 60, 90, 105, 106, 122, 126, 146, 150. \\
    & \\
    {\sf strong:} & 18, 22, 30, 45, 106, 122, 126, 146. \\
    {\sf moderate:} & - \\
    {\sf weak:} & 60, 90, 105, 150.
    \end{tabular}
\end{center}

\subsection{Communication complexity classification (2004)}

Communication complexity classification in ``Cellular Automata and Communication Complexity'', establishes three ECA classes:

\begin{center}
\{{\sf bounded, linear, other}\}.
\end{center}

For details please see \cite{kn:DRT04}.

\begin{center}
    \begin{tabular}{ r p{9cm}}
    {\bf bounded:} & 0, 1, 2, 3, 4, 5, 7, 8, 10, 12, 13, 15, 19, 24, 27, 28, 29, 32, 34, 36, 38, 42, 46, 51, 60, 71(29), 72, 76, 78, 90, 105, 108, 128, 130, 136, 138, 140, 150, 154, 156, 160, 162(missing), 170, 172, 200, 204. \\
    & \\
    {\sf strong:} & 2, 7, 10, 24, 34, 42, 46, 108, 128, 130, 138, 154, 162, 170. \\
    {\sf moderate:} & 1, 3, 4, 5, 8, 13, 15, 27, 28, 29, 32, 38, 72, 78, 136, 140, 156, 160, 172. \\
    {\sf weak:} & 0, 12, 19, 36, 51, 60, 76, 90, 105, 150, 200, 204.
    \end{tabular}
\end{center}

\begin{center}
    \begin{tabular}{ r p{9cm}}
    {\bf linear:} & 11, 14, 23, 33, 35, 43, 44, 50, 56, 58, 77, 132, 142, 152, 168, 178, 184, 232. \\
    & \\
    {\sf strong:} & 11, 35, 56, 58, 152, 178, 184. \\
    {\sf moderate:} & 14, 33, 43, 44, 77, 132, 142, 168. \\
    {\sf weak:} & 23, 50, 232.
    \end{tabular}
\end{center}

\begin{center}
    \begin{tabular}{ r p{9cm}}
    {\bf other:} & 6, 9, 18, 22, 25, 26, 30, 37, 40, 41, 45, 54, 57, 62, 73, 74, 94, 104, 106, 110, 122, 126, 134, 146, 164. \\
    & \\
    {\sf strong:} & 9, 18, 22, 25, 26, 30, 41, 45, 54, 57, 62, 94, 106, 110, 122, 126, 146. \\
    {\sf moderate:} & 6, 37, 40, 73, 74, 104, 134, 164. \\
    {\sf weak:} & -
    \end{tabular}
\end{center}

Additionally, bound class can be refined in other four subclasses.

\begin{center}
    \begin{tabular}{ r p{6cm}}
    {\bf bounded by additivity:} & 15, 51, 60, 90, 105, 108, 128, 136, 150, 160, 170, 204. \\
    & \\
    {\sf strong:} & 15, 51, 108, 128, 170. \\
    {\sf moderate:} & 136, 160. \\
    {\sf weak:} & 60, 90, 105, 150, 204.
    \end{tabular}
\end{center}

\begin{center}
    \begin{tabular}{ r p{4cm}}
    {\bf bounded by limited sensibility:} & 0, 1, 2, 3, 4, 5, 8, 10, 12, 19, 24, 29, 34, 36, 38, 42, 46, 72, 76, 78, 108, 138, 200. \\
    & \\
    {\sf strong:} & 2, 10, 24, 34, 42, 46, 108, 138. \\
    {\sf moderate:} & 1, 3, 4, 5, 8, 29, 38, 72, 78. \\
    {\sf weak:} & 0, 12, 19, 36, 76, 200.
    \end{tabular}
\end{center}

\begin{center}
    \begin{tabular}{ r p{6cm}}
    {\bf bounded by half-limited sensibility:} & 7, 13, 28, 140, 172. \\
    & \\
    {\sf strong:} & 7. \\
    {\sf moderate:} & 13, 28, 140, 172. \\
    {\sf weak:} & -
    \end{tabular}
\end{center}

\begin{center}
    \begin{tabular}{ r p{9cm}}
    {\bf bounded for any other reason:} & 27, 32, 130, 156, 162. \\
    & \\
    {\sf strong:} & 130, 162. \\
    {\sf moderate:} & 27, 32, 156. \\
    {\sf weak:} & -
    \end{tabular}
\end{center}

\subsection{Topological classification (2007)}

Topological classification in ``A Nonlinear Dynamics Perspective of Wolfram's New Kind of Science. Part VII: Isles of Eden'', establishes six ECA classes: 

\begin{center}
\{{\sf period-1, period-2, period-3, Bernoulli $\sigma_t$-shift, complex Bernoulli-shift, hyper Bernoully-shift}\}.
\end{center}

For details please see \cite{kn:CGS07}.

\begin{center}
    \begin{tabular}{ r p{9cm}}
    {\bf period-1:} & 0, 4, 8, 12, 13, 32, 36, 40, 44, 72, 76, 77, 78, 94, 104, 128, 132, 136, 140, 160, 164, 168, 172, 200, 204, 232. \\
    & \\
    {\sf strong:} & 94, 128. \\
    {\sf moderate:} & 4, 8, 13, 32, 40, 44, 72, 77, 78, 104, 132, 136, 140, 160, 164, 168, 172. \\
    {\sf weak:} & 0, 12, 36, 76, 200, 204, 232.
    \end{tabular}
\end{center}

\begin{center}
    \begin{tabular}{ r p{9cm}}
    {\bf period-2:} & 1, 5, 19, 23, 28, 29, 33, 37, 50, 51, 108, 156, 178. \\
    & \\
    {\sf strong:} & 108, 178. \\
    {\sf moderate:} & 1, 5, 28, 29, 33, 37, 156. \\
    {\sf weak:} & 19, 23, 50, 51.
    \end{tabular}
\end{center}

\begin{center}
    \begin{tabular}{ r p{9cm}}
    {\bf period-3:} & 62. \\
    & \\
    {\sf strong:} & 62. \\
    {\sf moderate:} & - \\
    {\sf weak:} & -
    \end{tabular}
\end{center}

\begin{center}
    \begin{tabular}{ r p{7cm}}
    {\bf Bernoulli $\sigma_t$-shift:} & 2, 3, 6 , 7, 9, 10, 11, 14, 15, 24, 25, 27, 34, 35, 38, 42, 43, 46, 56, 57, 58, 74, 130, 134, 138, 142, 152, 162, 170, 184. \\
    & \\
    {\sf strong:} & 2, 7, 9, 10, 11, 15, 24, 25, 34, 35, 42, 46, 56, 57, 58, 130, 138, 152, 162, 170, 184. \\
    {\sf moderate:} & 3, 6, 14, 27, 38, 43, 74, 134, 142. \\
    {\sf weak:} & -
    \end{tabular}
\end{center}

\begin{center}
    \begin{tabular}{ r p{9cm}}
    {\bf complex Bernoulli-shift:} & 18, 22, 54, 73, 90, 105, 122, 126, 146, 150. \\
    & \\
    {\sf strong:} & 18, 22, 122, 126, 146. \\
    {\sf moderate:} & 73. \\
    {\sf weak:} & 90, 105, 150.
    \end{tabular}
\end{center}

\begin{center}
    \begin{tabular}{ r p{9cm}}
    {\bf hyper Bernoully-shift:} & 26, 30, 41, 45, 60, 106, 110, 154. \\
    & \\
    {\sf strong:} & 26, 30, 41, 45, 110, 154. \\
    {\sf moderate:} & - \\
    {\sf weak:} & 60.
    \end{tabular}
\end{center}

\subsection{Power spectral classification (2008)}

Power spectral classification in ``Power Spectral Analysis of Elementary Cellular Automata'', establishes four ECA classes: 

\begin{center}
\{{\sf category 1: extremely low power density, category 2: broad-band noise, category 3: power law spectrum, exceptional rules}\}.
\end{center}

For details please see \cite{kn:Nina08}.

\begin{center}
    \begin{tabular}{ r p{4cm}}
    {\bf category 1  extremely low power density:} & 0, 1, 4, 5, 8, 12, 13, 19, 23, 26, 28, 29, 33, 37, 40, 44, 50, 51, 72, 76, 77, 78, 104, 128, 132, 133(94), 136, 140, 156, 160, 164, 168, 172, 178, 200, 232. \\
    & \\
    {\sf strong:} & 26, 94, 128, 178. \\
    {\sf moderate:} & 1, 4, 5, 8, 13, 28, 29, 33, 37, 40, 44, 72, 77, 78, 104, 132, 136, 140, 156, 160, 164, 168, 172. \\
    {\sf weak:} & 0, 12, 19, 23, 50, 51, 76, 200, 232.
    \end{tabular}
\end{center}

\begin{center}
    \begin{tabular}{ r p{6cm}}
    {\bf category 2 broad-band noise:} & 2, 3, 6, 7, 9, 10, 11, 14, 15, 18, 22, 24, 25, 26, 27, 30, 34, 35, 38, 41, 42, 43, 45, 46, 56, 57, 58, 60, 74, 90, 105, 106, 129(126), 130, 134, 138, 142, 146, 150, 152, 154, 161(122), 162, 170, 184. \\
    & \\
    {\sf strong:} & 2, 7, 9, 10, 11, 15, 18, 22, 24, 25, 26, 30, 34, 35, 41, 42, 45, 46, 56, 57, 58, 106, 122, 126, 130, 138, 146, 152, 154, 162, 170, 184. \\
    {\sf moderate:} & 3, 6, 14, 27, 38, 43, 74, 134, 142, . \\
    {\sf weak:} & 60, 90, 105, 150.
    \end{tabular}
\end{center}

\begin{center}
    \begin{tabular}{ r p{9cm}}
    {\bf category 3 power law spectrum:} & 54, 62, 110. \\
    & \\
    {\sf strong:} & 54, 62, 110. \\
    {\sf moderate:} & . \\
    {\sf weak:} & .
    \end{tabular}
\end{center}

\begin{center}
    \begin{tabular}{ r p{9cm}}
    {\bf exceptional rules:} & 73, 204. \\
    & \\
    {\sf strong:} & - \\
    {\sf moderate:} & 73. \\
    {\sf weak:} & 204.
    \end{tabular}
\end{center}

\subsection{Morphological diversity classification (2010)}

Morphological diversity classification in ``On Generative Morphological Diversity of Elementary Cellular Automata'', establishes five ECA classes:

\begin{center}
\{{\sf chaotic, complex, periodic, two-cycle, fixed point, null}\}.
\end{center}

For details please see \cite{kn:AM10}.

\begin{center}
    \begin{tabular}{ r p{9cm}}
    {\bf chaotic:} & 2, 10, 18, 22, 24, 26, 30, 34, 42, 45, 56, 60, 73, 74, 90, 94, 105, 106, 126, 130, 138, 150, 152, 154, 161(122), 162, 170, 184. \\
    & \\
    {\sf strong:} & 2, 10, 18, 22, 24, 26, 30, 34, 42, 56, 94, 106, 122, 126, 130, 138, 152, 154, 162, 170, 184. \\
    {\sf moderate:} & 73, 74. \\
    {\sf weak:} & 60, 90, 105, 150.
    \end{tabular}
\end{center}

\begin{center}
    \begin{tabular}{ r p{9cm}}
    {\bf complex:} & 54, 110. \\
    & \\
    {\sf strong:} & 54, 110. \\
    {\sf moderate:} & - \\
    {\sf weak:} & -
    \end{tabular}
\end{center}

\begin{center}
    \begin{tabular}{ r p{9cm}}
    {\bf periodic:} & 18, 26, 60, 90, 94, 154. \\
    & \\
    {\sf strong:} & 18, 26, 94, 154. \\
    {\sf moderate:} & - \\
    {\sf weak:} & 60, 90.
    \end{tabular}
\end{center}

\begin{center}
    \begin{tabular}{ r p{9cm}}
    {\bf two-cycle:} & 1, 2, 3, 4, 5, 6, 7, 9, 10, 11, 12, 13, 14, 15, 19, 23, 24, 25, 27, 28, 29, 33, 34, 35, 36, 37, 38, 42, 43, 44, 46, 50, 51, 56, 58, 74, 76, 106, 108, 130, 132, 134, 138, 140, 142, 152, 156, 162, 164, 170, 172, 178, 184, 204. \\
    & \\
    {\sf strong:} & 2, 7, 9, 10, 11, 15, 24, 25, 34, 35, 42, 46, 56, 58, 106, 108, 130, 138, 152, 162, 170, 178, 184. \\
    {\sf moderate:} & 1, 3, 4, 5, 6, 13, 14, 27, 28, 29, 33, 37, 38, 43, 44, 74, 132, 134, 140, 142, 156, 164, 172. \\
    {\sf weak:} & 12, 19, 23, 36, 50, 51, 76, 204.
    \end{tabular}
\end{center}

\begin{center}
    \begin{tabular}{ r p{9cm}}
    {\bf fixed point:} & 0, 2, 4, 8, 10, 11, 12, 13, 14, 24, 32, 34, 36, 40, 42, 43, 44, 46, 50, 56, 57, 58, 72, 74, 76, 77, 78, 104, 106, 108, 128, 130, 132, 136, 138, 140, 142, 152, 160, 162, 164, 168, 170, 172, 178, 184, 200, 204, 232. \\
    & \\
    {\sf strong:} & 2, 10, 11, 24, 34, 42, 46, 56, 57, 58, 106, 108, 128, 130, 138, 152, 162, 170, 178, 184. \\
    {\sf moderate:} & 4, 8, 13, 14, 32, 40, 43, 44, 72, 74, 77, 78, 104, 132, 136, 140, 142, 160, 164, 168, 172. \\
    {\sf weak:} & 0, 12, 36, 50, 76, 200, 204, 232.
    \end{tabular}
\end{center}

\begin{center}
    \begin{tabular}{ r p{9cm}}
    {\bf null:} & 0, 8, 32, 40, 72, 104, 128, 136, 160, 168, 200, 232. \\
    & \\
    {\sf strong:} & 128. \\
    {\sf moderate:} & 8, 32, 40, 72, 104, 136, 160, 168. \\
    {\sf weak:} & 0, 200, 232.
    \end{tabular}
\end{center}

\subsection{Distributive and non-distributive lattices classification (2010)}

Distributive and non-distributive lattices classification in ``Inducing Class 4 Behavior on the Basis of Lattice Analysis'', establishes four ECA classes:

\begin{center}
\{{\sf class 1, class 2, class 3, class 4}\}.
\end{center}

For details please see \cite{kn:Gunji10}.

\begin{center}
    \begin{tabular}{ r p{9cm}}
    {\bf class 1:} & 0, 32, 128, 160, 250(160), 254(128). \\
    & \\
    {\sf strong:} & 128. \\
    {\sf moderate:} & 32, 160. \\
    {\sf weak:} & 0.
    \end{tabular}
\end{center}

\begin{center}
    \begin{tabular}{ r p{9cm}}
    {\bf class 2:} & 4, 36, 50, 72, 76, 94, 104, 108, 132, 164, 178, 200, 204, 218(164), 232, 236(200). \\
    & \\
    {\sf strong:} & 94, 108, 178. \\
    {\sf moderate:} & 4, 72, 104, 132, 164. \\
    {\sf weak:} & 36, 50, 76, 200, 204, 232.
    \end{tabular}
\end{center}

\begin{center}
    \begin{tabular}{ r p{9cm}}
    {\bf class 3:} & 18, 22, 54, 122, 126, 146, 150, 182(146). \\
    & \\
    {\sf strong:} & 18, 22, 54, 122, 126, 146. \\
    {\sf moderate:} & - \\
    {\sf weak:} & 150.
    \end{tabular}
\end{center}

\begin{center}
    \begin{tabular}{ r p{9cm}}
    {\bf class 4:} & 110, 124(110), 137(110), 193(110). \\
    & \\
    {\sf strong:} & 110. \\
    {\sf moderate:} & . \\
    {\sf weak:} & .
    \end{tabular}
\end{center}

\subsection{Topological dynamics classification (2012)}

Topological classification in ``A Full Computation-Relevant Topological Dynamics Classification of Elementary Cellular Automata'', establishes four ECA classes: 

\begin{center}
\{{\sf equicontinuous, almost-equicontinuous, sensitive, sensitive positively expansive}\}.
\end{center}

For details please see \cite{kn:SS12, kn:CFM00}.

\begin{center}
    \begin{tabular}{ r p{9cm}}
    {\bf equicontinuous:} & 0, 1, 4, 5, 8, 12, 19, 29, 36, 51, 72, 76, 108, 200, 204. \\
    & \\
    {\sf strong:} & 108. \\
    {\sf moderate:} & 1, 4, 5, 8, 29, 72. \\
    {\sf weak:} & 0, 12, 19, 36, 51, 76, 200, 204.
    \end{tabular}
\end{center}

\begin{center}
    \begin{tabular}{ r p{6cm}}
    {\bf almost-equicontinuous:} & 13, 23, 28, 32, 33, 40, 44, 50, 73, 77, 78, 94, 104, 128, 132, 136, 140, 156, 160, 164, 168, 172, 178, 232. \\
    & \\
    {\sf strong:} & 94, 128, 178. \\
    {\sf moderate:} & 13, 28, 32, 40, 73, 77, 78, 104, 132, 136, 140, 156, 160, 164, 168, 172. \\
    {\sf weak:} & 23, 50, 232.
    \end{tabular}
\end{center}

\begin{center}
    \begin{tabular}{ r p{9cm}}
    {\bf sensitive:} & 2, 3, 6, 7, 9, 10, 11, 14, 15, 18, 22, 24, 25, 26, 27, 30, 34, 35, 37, 38, 41, 42, 43, 45, 46, 54, 56, 57, 58, 60, 62, 74, 106, 110, 122, 126, 130, 134, 138, 142, 146, 152, 154, 162, 170, 184. \\
    & \\
    {\sf strong:} & 2, 7, 9, 10, 11, 15, 18, 22, 24, 25, 26, 30, 34, 35, 41, 42, 45, 46, 54, 56, 57, 58, 62, 106, 110, 122, 126, 130, 138, 146, 152, 154, 162, 170, 184. \\
    {\sf moderate:} & 3, 6, 14, 27, 37, 38, 43, 74, 134, 142. \\
    {\sf weak:} & 60.
    \end{tabular}
\end{center}

\begin{center}
    \begin{tabular}{ r p{9cm}}
    {\bf sensitive positively expansive:} & 90, 105, 150. \\
    & \\
    {\sf strong:} & - \\
    {\sf moderate:} & - \\
    {\sf weak:} & 90, 105, 150.
    \end{tabular}
\end{center}

Also, this classification can be refined into three sub-classes: weakly periodic, surjective, and chaotic (in the sense of Denavey).

\begin{center}
    \begin{tabular}{ r p{9cm}}
    {\bf weakly periodic:} & 2, 3, 10, 15, 24, 34, 38, 42, 46, 138, 170. \\
    & \\
    {\sf strong:} & 2, 10, 15, 24, 34, 42, 46, 138, 170. \\
    {\sf moderate:} & 3, 38. \\
    {\sf weak:} & -
    \end{tabular}
\end{center}

\begin{center}
    \begin{tabular}{ r p{9cm}}
    {\bf surjective:} & 15, 30, 45, 51, 60, 90, 105, 106, 150, 154, 170, 204. \\
    & \\
    {\sf strong:} & 15, 30, 45, 154, 170. \\
    {\sf moderate:} & -  \\
    {\sf weak:} & 51, 60, 90, 105, 150, 204.
    \end{tabular}
\end{center}

\begin{center}
    \begin{tabular}{ r p{5cm}}
    {\bf chaotic (in the sense of Denavey):} & 15, 30, 45, 60, 90, 105, 106, 150, 154, 170. \\
    & \\
    {\sf strong:} & 15, 30, 45, 106, 154, 170. \\
    {\sf moderate:} & - \\
    {\sf weak:} & 60, 90, 105, 150.
    \end{tabular}
\end{center}

\subsection{Expressivity analysis (2013)}

This is a classification by the evolution of a configuration consisting of an isolated one surrounded by zeros, that is a bit different from conventional ECA classifications previously displayed. In ``Expressiveness of Elementary Cellular Automata'', we can see five ECA kinds of expressivity:

\begin{center}
\{{\sf 0, periodic patterns, complex, Sierpinski patterns, finite growth}\}.
\end{center}

For details please see \cite{kn:RAM13}.

\begin{center}
    \begin{tabular}{ r p{9cm}}
    {\bf 0:} & 0, 7, 8, 19, 23, 31, 32, 40, 55, 63, 72, 104, 127, 128, 136, 160, 168, 200, 232. \\
    & \\
    {\sf strong:} & 7, 128. \\
    {\sf moderate:} & 8, 32, 40, 72, 104, 136, 160, 168. \\
    {\sf weak:} & 0, 19, 23, 200, 232.
    \end{tabular}
\end{center}

\begin{center}
    \begin{tabular}{ r p{8cm}}
    {\bf periodic patterns:} & 13, 28, 50, 54, 57, 58, 62, 77, 78, 94, 99, 109, 122, 156, 178. \\
    & \\
    {\sf strong:} & 54, 57, 58, 62, 94, 122, 178. \\
    {\sf moderate:} & 13, 28, 73, 77, 78, 156. \\
    {\sf weak:} & 50.
    \end{tabular}
\end{center}

\begin{center}
    \begin{tabular}{ r p{9cm}}
    {\bf complex:} & 30, 45, 73, 75, 110. \\
    & \\
    {\sf strong:} & 30, 45, 110. \\
    {\sf moderate:} & 73. \\
    {\sf weak:} & -
    \end{tabular}
\end{center}

\begin{center}
    \begin{tabular}{ r p{9cm}}
    {\bf Sierpinski patterns:} & 18, 22, 26, 60, 90, 105, 126, 146, 150, 154. \\
    & \\
    {\sf strong:} & 18, 22, 26, 126, 146, 154. \\
    {\sf moderate:} & - \\
    {\sf weak:} & 60, 90, 105, 150.
    \end{tabular}
\end{center}

\begin{center}
    \begin{tabular}{ r p{9cm}}
    {\bf finite growth:} & 1, 2, 3, 4, 5, 6, 9, 10, 11, 12, 14, 15, 24, 25, 27, 29, 33, 34, 35, 36, 37, 38, 39, 41, 42, 43, 44, 46, 47, 51, 56, 59, 71, 74, 76, 103, 106, 107, 108, 111, 130, 132, 134, 138, 140, 142, 152, 162, 164, 170, 172, 184, 204. \\
    & \\
    {\sf strong:} & 2, 9, 10, 11, 15, 24, 25, 34, 35, 41, 42, 46, 56, 106, 108, 130, 152, 162, 170, 184. \\
    {\sf moderate:} & 1, 3, 4, 5, 6, 14, 27, 29, 33, 37, 38, 43, 44, 74, 140, 142, 164, 172. \\
    {\sf weak:} & 12, 36, 51, 76, 204.
    \end{tabular}
\end{center}

\subsection{Normalised compression classification (2013)}

Normalised compression classification in ``Asymptotic Behaviour and Ratios of Complexity in Cellular Automata Rule Spaces'', establishes two ECA classes:

\begin{center}
\{{\sf $C_{1,2}$, $C_{3,4}$}\}.
\end{center}

For details please see \cite{kn:ZZ}.

\begin{center}
    \begin{tabular}{ r p{9cm}}
    {\bf $C_{1,2}$:} & 0, 1, 2, 3, 4, 5, 6, 7, 8, 9, 10, 11, 12, 13, 14, 15, 19, 23, 24, 25, 26, 27, 28, 29, 32, 33, 34, 35, 36, 37, 38, 40, 42, 43, 44, 46, 50, 51, 56, 57, 58, 72, 74, 76, 77, 78, 104, 108, 128, 130, 132, 134, 136, 138, 140, 142, 152, 154, 156, 160, 162, 164, 168, 170, 172, 178, 184, 200, 204, 232. \\
    & \\
    {\sf strong:} & 2, 7, 9, 10, 11, 15, 24, 25, 26, 34, 35, 42, 46, 56, 57, 58, 108, 128, 130, 138, 152, 154, 170, 178, 184. \\
    {\sf moderate:} & 1, 3, 4, 5, 6, 8, 13, 14, 27, 28, 29, 32, 33, 37, 38, 40, 43, 44, 72, 74, 77, 78, 104, 132, 134, 136, 140, 142, 156, 160. \\
    {\sf weak:} & 0, 12, 19, 23, 36, 50, 51, 76, 200, 204, 232.
    \end{tabular}
\end{center}

\begin{center}
    \begin{tabular}{ r p{9cm}}
    {\bf $C_{3,4}$:} & 18, 22, 30, 41, 45, 54, 60, 62, 73, 90, 94, 105, 106, 110, 122, 126, 146, 150. \\
    & \\
    {\sf strong:} & 18, 22, 30, 41, 45, 54, 62, 94, 106, 110, 122, 126, 146. \\
    {\sf moderate:} & 73. \\
    {\sf weak:} & 60, 90, 105, 150.
    \end{tabular}
\end{center}

\subsection{Surface dynamics classification (2013)}

Expressivity classification in ``Emergence of Surface Dynamics in Elementary Cellular Automata'', establishes three ECA classes: 

\begin{center}
\{{\sf type A, type B, type C}\}.
\end{center}

For details please see \cite{kn:MMM}.

\begin{center}
    \begin{tabular}{ r p{9cm}}
    {\bf type A:} & 0, 1, 2, 3, 4, 5, 6, 7, 8, 9, 10, 11, 12, 13, 14, 15, 19, 23, 24, 25, 27, 28, 29, 32, 33, 34, 35, 36, 37, 38, 40, 42, 43, 44, 46, 50, 51, 56, 57, 58, 72, 74, 76, 77, 78, 94, 104, 108, 128, 130, 132, 134, 136, 138, 140, 142, 152, 156, 160, 162, 164, 168, 170, 172, 178, 184, 200, 204, 232. \\
    & \\
    {\sf strong:} & 2, 7, 9, 10, 11, 15, 24, 25, 34, 35, 42, 46, 56, 57, 58, 128, 130, 152, 170, 178, 184. \\
    {\sf moderate:} & 1, 3, 4, 5, 6, 8, 13, 14, 27, 28, 29, 32, 33, 37, 38, 40, 43, 44, 72, 74, 77, 78, 104, 108, 132, 134, 136, 140, 142, 156, 160, 164, 168, 172. \\
    {\sf weak:} & 0, 12, 19, 23, 36, 50, 51, 76, 200, 204, 232.
    \end{tabular}
\end{center}

\begin{center}
    \begin{tabular}{ r p{9cm}}
    {\bf type B:} & 18, 22, 26, 30, 41, 45, 60, 90, 105, 106, 122, 126, 146, 150, 154. \\
    & \\
    {\sf strong:} & 18, 22, 26, 30, 45, 106, 122, 126, 146, 154. \\
    {\sf moderate:} & - \\
    {\sf weak:} & 18, 22, 26, 30, 45, 106, 122, 126, 146, 154.
    \end{tabular}
\end{center}

\begin{center}
    \begin{tabular}{ r p{9cm}}
    {\bf type C:} & 54, 62, 73, 110. \\
    & \\
    {\sf strong:} & 54, 62, 110. \\
    {\sf moderate:} & 73. \\
    {\sf weak:} & -
    \end{tabular}
\end{center}

\subsection{Spectral classification (2013)}

Spectral classification in ``A Spectral Portrait of the Elementary Cellular Automata Rule Space'', establishes four ECA classes: 

\begin{center}
\{{\sf DE/SFC, DE/SFC SFC, EB, S}\}.
\end{center}

For details please see \cite{kn:RO13}.

\begin{center}
    \begin{tabular}{ r p{9cm}}
    {\bf DE/SFC:} & 0, 1, 2, 5, 6, 7, 8, 9, 10, 11, 12, 14, 19, 22, 23, 24, 25, 26, 27, 29, 32, 33, 34, 35, 36, 37, 38, 40, 41, 42, 43, 44, 46, 50, 54, 56, 57, 58, 62, 72, 73, 74, 76, 77, 94, 104, 108, 110, 128, 130, 132, 134, 136, 138, 140, 142, 152, 160, 162, 164, 168, 172, 178, 184, 200, 232. \\
    & \\
    {\sf strong:} & 2, 7, 9, 10, 11, 22, 24, 25, 26, 29, 34, 35, 41, 42, 46, 54, 56, 57, 58, 62, 94, 108, 110, 128, 130, 138, 152, 162, 178, 184. \\
    {\sf moderate:} & 1, 5, 6, 8, 14, 27, 32, 33, 37, 38, 40, 43, 44, 72, 73, 74, 77, 104, 132, 134, 136, 140, 142, 160, 164, 168, 172. \\
    {\sf weak:} & 0, 12, 19, 23, 36, 50, 76, 200, 232.
    \end{tabular}
\end{center}

\begin{center}
    \begin{tabular}{ r p{9cm}}
    {\bf DE/SFC SFC:} & 3, 4. \\
    & \\
    {\sf strong:} & - \\
    {\sf moderate:} & 3, 4. \\
    {\sf weak:} & -
    \end{tabular}
\end{center}

\begin{center}
    \begin{tabular}{ r p{9cm}}
    {\bf EB:} & 13, 18, 28, 78, 122, 126, 146, 156. \\
    & \\
    {\sf strong:} & 18, 122, 126, 146. \\
    {\sf moderate:} & 13, 28, 78, 156. \\
    {\sf weak:} & -
    \end{tabular}
\end{center}

\begin{center}
    \begin{tabular}{ r p{9cm}}
    {\bf S:} & 15, 30, 45, 51, 60, 90, 105, 106, 150, 154, 170, 204. \\
    & \\
    {\sf strong:} & 15, 30, 45, 106, 154, 170. \\
    {\sf moderate:} & - \\
    {\sf weak:} & 51, 60, 90, 105, 150, 204.
    \end{tabular}
\end{center}

\subsection{Bijective and surjective classification (2013)}

In this section, we have just bijective and surjective classification (personal communication, Harold V. McIntosh and Juan C. Seck Tuoh Mora): 

\begin{center}
\{{\sf bijective, surjective}\}.
\end{center}

For details please see \cite{kn:Mc90, kn:Mc09}.

\begin{center}
    \begin{tabular}{ r p{9cm}}
    {\bf bijective:} & 15, 51, 170, 204. \\
    & \\
    {\sf strong:} & 15, 170. \\
    {\sf moderate:} & - \\
    {\sf weak:} & 51, 204.
    \end{tabular}
\end{center}

\begin{center}
    \begin{tabular}{ r p{9cm}}
    {\bf surjective:} & 30, 45, 60, 90, 105, 106, 150, 154. \\
    & \\
    {\sf strong:} & 30, 45, 106, 154. \\
    {\sf moderate:} & - \\
    {\sf weak:} & 60, 90, 105, 150.
    \end{tabular}
\end{center}

\subsection{Creativity classification (2013)}

Creativity classification in ``On Creativity of Elementary Cellular Automata'', establishes four ECA classes: 

\begin{center}
\{{\sf creative, schizophrenic, autistic savants, severely autistic}\}.
\end{center}

For details please see \cite{kn:AW}.

\begin{center}
    \begin{tabular}{ r p{9cm}}
    {\bf creative:} & 3, 5, 11, 13, 15, 35. \\
    & \\
    {\sf strong:} & 11, 15, 35. \\
    {\sf moderate:} & 3, 5, 13. \\
    {\sf weak:} & -
    \end{tabular}
\end{center}

\begin{center}
    \begin{tabular}{ r p{9cm}}
    {\bf schizophrenic:} & 9, 18, 22, 25, 26, 28, 30, 37, 41, 43, 45, 54, 57, 60, 62, 73, 77, 78, 90, 94, 105, 110, 122, 126, 146, 150, 154, 156. \\
    & \\
    {\sf strong:} & 9, 18, 22, 25, 26, 30, 41, 45, 54, 57, 62, 110, 122, 126, 146, 152, 154. \\
    {\sf moderate:} & 28, 37, 43, 73, 77, 78, 156. \\
    {\sf weak:} & 60, 90, 105.
    \end{tabular}
\end{center}

\begin{center}
    \begin{tabular}{ r p{8cm}}
    {\bf autistic savants:} & 1, 2, 4, 7, 8, 10, 12, 14, 19, 32, 34, 42, 50, 51, 76, 128, 136, 138, 140, 160, 162, 168, 170, 200, 204. \\
    & \\
    {\sf strong:} & 2, 7, 10, 34, 42, 128, 138, 162, 170. \\
    {\sf moderate:} & 1, 4, 8, 14, 32, 136, 140, 160, 168. \\
    {\sf weak:} & 12, 19, 50, 51, 76, 200, 204.
    \end{tabular}
\end{center}

\begin{center}
    \begin{tabular}{ r p{8cm}}
    {\bf severely autistic:} & 23, 24, 27, 29, 33, 36, 40, 44, 46, 56, 58, 72, 74, 104, 106, 108, 130, 132, 142, 152, 164, 172, 178, 184, 232. \\
    & \\
    {\sf strong:} & 24, 46, 56, 58, 106, 108, 130, 152, 178, 184. \\
    {\sf moderate:} & 27, 29, 33, 40, 44, 72, 74, 104, 132, 142, 164, 172. \\
    {\sf weak:} & 23, 36, 232.
    \end{tabular}
\end{center}

\section*{Acknowledgement}
Thanks to several authors of all previous classifications for discussion and improve this note.


\end{document}